\documentclass[fleqn,usenatbib]{mnras}

\usepackage{newtxtext,newtxmath}
\usepackage[T1]{fontenc}

\DeclareRobustCommand{\VAN}[3]{#2}
\let\VANthebibliography\thebibliography
\def\thebibliography{\DeclareRobustCommand{\VAN}[3]{##3}\VANthebibliography}

\usepackage{graphicx}	% Including figure files
\usepackage{amsmath}	% Advanced maths commands
\usepackage{inputenc}
\usepackage{color, natbib, color} %apjfonts,
\usepackage{multirow, placeins, subfigure}
\usepackage{epstopdf}
\usepackage{bm}
\usepackage{overpic}
\usepackage{placeins}
\usepackage{afterpage}
\usepackage{hyperref}
\usepackage{xcolor}
\newcommand\gradunit{\mathrm{dex\,kpc^{-1}}}
\newif\ifshowcomt
\newif\ifshowmaintext
\showcomtfalse  % 若要取消注释，将此行改为 \showcomtfalse
\showmaintexttrue
\newcommand{\comt}[1]{\ifshowcomt [\textit{#1}] \fi}

\newcommand{\annotatedtext}[2]{%
  \comt{#1}%
  \ifshowmaintext
    #2%
  \fi
}

\newif\ifusemyemph
\usemyemphfalse % 取消强调效果（只显示原始文本）

\newcommand{\myemph}[1]{%
  \ifusemyemph
    \textbf{\uline{#1}}%
  \else
    #1%
  \fi
}

\newif\ifusenewemph
\usenewemphfalse % 取消强调效果（只显示原始文本）

\newcommand{\newemph}[1]{%
  \ifusenewemph
    \textbf{#1}%
  \else
    #1%
  \fi
}
\newcommand{\triemph}[1]{%
  \ifusetriemph
    \textbf{#1}%
  \else
    #1%
  \fi
}
\newif\ifusetriemph
\usetriemphfalse % 取消强调效果（只显示原始文本）

\newcommand{\revise}[1]{%
  \ifuserevise
    \textbf{#1}%
  \else
    #1%
  \fi
}
\newif\ifuserevise
\userevisefalse % 取消强调效果（只显示原始文本）
\title[Metallicity Gradient-Age Relation in the MW]{Revisiting the Radial Metallicity Gradient-Age Relation in the Milky Way's Thin and Thick Disks}

\author[Ao Chen et al.]{
Ao Chen,$^{1,2,3}$
Juntai Shen,$^{1,2,3}$\thanks{E-mail: jtshen@sjtu.edu.cn; wchun@tjnu.edu.cn; huangyang@ucas.ac.cn}
Chun Wang,$^{4}$
and Yang Huang$^{5,6}$
\\
$^{1}$Department of Astronomy, School of Physics and Astronomy, Shanghai Jiao Tong University, 800 Dongchuan Road, Shanghai 200240, PRC\\
$^{2}$State Key Laboratory of Dark Matter Physics, School of Physics and Astronomy, Shanghai Jiao Tong University, Shanghai 200240, PRC\\
$^{3}$Key Laboratory for Particle Astrophysics and Cosmology (MOE)/Shanghai Key Laboratory for Particle Physics and Cosmology, Shanghai 200240, PRC\\
$^{4}$Tianjin Astrophysics Center, Tianjin Normal University, Tianjin 300387, PRC\\
$^{5}$School of Astronomy and Space Science, University of Chinese Academy of Sciences, Beijing, PRC\\
$^{6}$New Cornerstone Science Laboratory, National Astronomical Observatories, Chinese Academy of Sciences, Beijing, PRC
}

\date{Accepted XXX. Received YYY; in original form ZZZ}

\pubyear{2025}

\begin{document}
\label{firstpage}
\pagerange{\pageref{firstpage}--\pageref{lastpage}}
\maketitle

\begin{abstract}
Galactic disks typically exhibit a negative radial metallicity gradient, indicating faster enrichment in the inner regions. Recent studies report that this gradient becomes flatter with increasing stellar age in the Milky Way’s (MW) thin disk, while the thick disk exhibits a mildly positive gradient across all ages. In this work, we revisit the metallicity gradient–age relation (MGAR) in both the thin and thick disks of the MW, using spectroscopic data from LAMOST DR8 and stellar ages calibrated with asteroseismology. Our results show a steadily flattening MGAR in the thin disk and confirm a positive gradient $\sim0.013\,\gradunit$ in the thick disk. The flattening in the thin disk may be caused by large-scale radial migration induced by transient spiral arms, or by a time-dependent steepening of the interstellar medium (ISM) metallicity gradient as suggested by recent FIRE2 simulations. The positive gradient in the thick disk may reflect early enrichment of the outer regions by strong feedback or starburst-driven outflows in a turbulent, gas-rich proto-disk. These findings suggest distinct chemodynamical evolution paths for the MW's thin and thick disks and provide valuable constraints for future models of Galactic chemical evolution.
\end{abstract}


%%%%%%%%%%%%%%%%%%%%%%%%%%%%%%%%%%%%%%%%%%%%%%%%%%

%%%%%%%%%%%%%%%%% BODY OF PAPER %%%%%%%%%%%%%%%%%%

\section{Introduction}\label{section:introduction}
A cornerstone of galaxy evolution is the negative radial metallicity gradient, which describes the decrease in stellar metallicity with increasing galactocentric radius. \citet{1989MNRAS.239..885M} reproduced this trend using “inside-out” formation models, in which the inner disk forms stars more rapidly and enriches earlier than the outer regions. This scenario naturally results in a negative metallicity gradient for stars formed from the interstellar medium (ISM). 

Nowadays, the radial metallicity gradient has become a crucial diagnostic in different classes of models of Galactic evolution. First, chemical evolution models incorporate physical processes such as radial gas flows, star formation histories, and stellar yields to get a realistic metallicity gradient and study the formation of the MW (e.g., \citealt{2018MNRAS.481.2570G, 2021MNRAS.502.5935S}; see \citealt{2021A&ARv..29....5M} for a review). Second, descriptive models treat the gradient as a parameter to fit observational data and constrain the structure of the Galactic disk \citep[e.g.,][]{2009MNRAS.396..203S, 2015MNRAS.449.3479S, 2020ApJ...896...15F, 2021MNRAS.507.5882S}. Finally, hybrid models combine chemical evolution with numerical simulations, as in \citet{2013A&A...558A...9M} and \citet{2021MNRAS.508.4484J}. Utilizing the hybrid framework of \citet{2013A&A...558A...9M}, \citet{2018MNRAS.481.1645M} \revise{inferred stellar birth radius and} concluded that the ISM gradient flattens over time as the outer disk becomes progressively enriched and approaches the central metallicity.

In this context, we refer to the \textit{stellar birth gradient} as the radial metallicity gradient imprinted on stars at their formation, which inherits from the corresponding ISM gradient. Reconstructing the time sequence of these birth gradients would require direct knowledge of the metallicity distribution of newly formed stars at each epoch, which is not \triemph{directly observable}. Instead, what we can measure today is the metallicity gradient of mono-age stellar populations. By using stellar age as a proxy for look-back time, these present-day gradients give rise to the \textit{metallicity gradient–age relation (MGAR)}. It is also important to note that the present-day radial distribution of a mono-age stellar population does not necessarily preserve the distribution of their birth radii, as stars can migrate significantly over their lifetimes. Consequently, the observed MGAR is shaped by both the time evolution of the stellar birth gradient and the subsequent dynamical history of the Milky Way (MW).

\annotatedtext{Early (equivalent) MGAR observations.}{The pioneering study on the MGAR in our MW by \citet{1976A&A....48..301M} measured a mean radial metallicity gradient of $-0.05\pm 0.01\,\gradunit$, which steepens to $-0.10\pm 0.02\,\gradunit$ for younger stars. Using effective temperature ($T_\mathrm{eff}$) of main-sequence stars as a proxy for age, \citet{2012ApJ...754..124Y} found that the gradient steepens with increasing $T_\mathrm{eff}$ (typically associated with decreasing age), but shows no trend for stars with $T_\mathrm{eff} < 6,000\,\mathrm{K}$ (i.e., those with main-sequence lifetimes $\gtrsim 10$ Gyr). These early findings hinted at a flattening gradient for increasingly older stars in the MW's thin disk.} %This implies a distinct MGAR between the thin and thick disks, as the thin disk's age is younger than $10$ Gyr.}

\annotatedtext{Modern MGAR observations.}{Large-scale spectroscopic surveys have greatly facilitated precise age estimates, enabling direct MGAR measurements \citep[e.g.,][]{2021ApJ...922..189V, 2023A&A...674A.129W, 2023MNRAS.526.2141W, 2023MNRAS.525.2208R}. \citet[hereafter V21]{2021ApJ...922..189V} used LAMOST DR5 and reported a flattening gradient in the thin disk, starting at $-0.075\,\gradunit$ for the youngest stars and flattening with a rate of $0.003\,\mathrm{dex\, kpc^{-1}\, Gyr^{-1}}$. \citet{2023MNRAS.526.2141W} used asteroseismic age and adopted a hierarchical Bayesian model for the gradient fitting. They found a similar flattening trend as V21 in the thin disk.} In the thick disk, V21 found a slightly positive gradient of $\sim0.02\, \gradunit$ across all age bins, consistent with constantly positive gradients of old, cool main-sequence stars ($T_\mathrm{eff}<6,000\,\mathrm{K}$) in \citet{2012ApJ...754..124Y}, which are likely from the thick disk.

\annotatedtext{Introduce churning and blurring.}{Radial migration has been proposed as a major process that alters stellar positions and flattens metallicity gradients \citep{2002MNRAS.336..785S}. This mechanism, sometimes dubbed ``churning'', causes significant radial mixing; non-axisymmetric perturbations such as transient spiral arms can induce large-scale stellar migration by exchanging stars across the corotation radius. In contrast, ``blurring'' refers to orbital heating that increases the amplitude of stellar epicycles without changing their guiding centers. Churning flattens the MGAR as older stars have had more time to migrate, resulting in a flatter radial metallicity distribution.}

\revise{The metallicity gradients at stellar birth provided by the recent method\footnote{\url{https://github.com/BridgetRatcliffe/Rbirth}} for estimating the stellar birth radius ($R_\mathrm{birth}$) further supports the significance of radial migration. Building on the empirical framework of \citet{2018MNRAS.481.1645M}, a more systematic approach to infer $R_\mathrm{birth}$ from stellar ages and metallicities has been developed in the follow-up studies \citep{2024MNRAS.535..392L, 2023MNRAS.525.2208R, 2025arXiv250902691R, 2025A&A...698A.267R}. In these works, the evolution of the stellar birth metallicity gradient is approximated as}
\begin{equation}
\nabla\mathrm{[Fe/H]}(\tau)\approx 
\frac{\Delta \mathrm{[Fe/H]}(\tau)}{\Delta R(\tau)}
\propto 
\frac{\mathrm{Range[Fe/H](age)}}{\Delta R(\tau)},
\end{equation}
\revise{where $\tau$ is the lookback time and equals stellar age. A linear anti-correlation between $\nabla\mathrm{[Fe/H]}(\tau)$ and $\mathrm{Range[Fe/H](age)}$---validated with NIHAO-UHD \citep{2020MNRAS.491.3461B} and HESTIA \citep{2020MNRAS.498.2968L} simulations by \citet{2024MNRAS.535..392L}---then provides an estimate of the birth gradient $\nabla\mathrm{[Fe/H]}(\tau)$ (with later refinements correcting $\Delta R(\tau)$ and accounting for selection effects; see \citealt{2025arXiv250902691R, 2025A&A...698A.267R}). Finally, the birth radius is obtained as}
\begin{equation}
R_\mathrm{birth}(\mathrm{age}, \mathrm{[Fe/H]})=
\frac{\mathrm{[Fe/H]}-\mathrm{[Fe/H]}(R=0,\tau)}
{\nabla\mathrm{[Fe/H]}(\tau)},
\end{equation}
\revise{given an estimate of $\mathrm{[Fe/H]}(R=0, \tau)$.}

\revise{As a by-product of this method, the reconstructed stellar birth gradients $\nabla\mathrm{[Fe/H]}(\tau)$ become progressively flatter with decreasing lookback time/age. \citet{2023A&A...678A.158A} further compared APOGEE MGAR measurements with the flattening birth-gradient evolution of \citet{2023MNRAS.525.2208R}, demonstrating how radial migration could explain the difference between  birth gradient evolution and observed MGAR trend.}

\revise{On the simulation side, models with smooth/bursty stellar feedback predict qualitatively different birth-gradient evolutions with redshift. As summarized in \citet{2025ApJ...989..147G}, simulations with \emph{smooth feedback} (such as EAGLE, Illustris, IllustrisTNG, and SIMBA) employ subgrid treatments of the ISM that produce relatively smooth metal redistribution and predict strong negative gradients \emph{flattening} with time. In contrast, \emph{bursty-feedback} simulations (e.g., FIRE2, SPICE Bursty, and Thesan Zoom) yield relatively \emph{shallower} birth gradients at all epochs and better match current observations at high redshifts \citep{2025arXiv251026877G}.}

\revise{In particular, recent FIRE2 simulations strongly challenge the migration-based interpretation of the MGAR \citep{2024arXiv241021377G, 2025ApJ...981...47G}. In these models, the present-day MGAR closely tracks the stellar birth gradients, implying minimal stellar radial migration in the simulations. And as an alternative to \textit{stellar} migration, \textit{gas} mixing (diffusion, advection, and turbulence) efficiently mixes gas-phase metals before forming new stars.}

\revise{Moreover, \citet{2024arXiv241021377G} argued that vertical evolution, where turbulent gas gradually settles over time (“upside-down” disk formation), may play a more important role than inside-out radial growth in shaping the Galaxy’s chemo-dynamical history. This highlights the need to account for vertical processes and potentially distinct evolutionary pathways of the thin and thick disks, as also emphasized by V21.}

\annotatedtext{Describe this work.}{To constrain the MGAR in the MW’s thin and thick disks, we follow an approach similar to V21. We isolate “pure” thin and thick disk populations using both geometric and chemical criteria, and compute MGAR separately for each population. Stellar kinematics and abundances are taken from LAMOST DR8 \citep{2022ApJS..259...51W}, and stellar ages are calibrated from asteroseismic data provided by the \textit{Kepler} mission \citep{2010PASP..122..131G,  2010Sci...327..977B, 2023A&A...675A..26W}, offering an alternative to traditional isochrone fitting.}

\annotatedtext{Paper structure.}{This paper is organized as follows. Section \ref{section:Data and method} describes the data selection, disk classification, orbital calculation, and gradient fitting method. Section \ref{section:Metallicity Gradient} presents the MGAR measurements in both the thin and thick disks. Section \ref{section:discussion} discusses the implications of our results and compares MGAR in the thin disk to the literature. Section \ref{section:conclusion} summarizes our conclusions.}

\section{Data and Method}\label{section:Data and method}

\annotatedtext{Dataset description.}{The Large Area Multi-Object Spectroscopic Telescope \citep[LAMOST, ][]{2012RAA....12..735D, 2012RAA....12..723Z, 2014IAUS..298..310L}, located at Xinglong Station of the National Astronomical Observatory of China, is a 4-meter Schmidt telescope with a 5° field of view. It is one of the most extensive stellar spectroscopic surveys and has released 13 data sets to date. In this work, we use LAMOST Data Release 8 \citep[DR8;][]{2022ApJS..259...51W} combined with stellar ages from \citet{2023A&A...675A..26W}, which trained a neural network using asteroseismic data from the LAMOST-\textit{Kepler} common stars.

Our cross-matched sample comprises 986,405 unique stars, including 688,369 red giant branch (RGB) stars and 298,036 red clump (RC) stars, each with 6D phase-space coordinates, [Fe/H], [$\alpha$/Fe], and age measurements. This study treats RGB and RC stars as a single population. We have also tested the MGAR trends separately for RGB and RC stars and found that the qualitative trends remain unchanged.}

\subsection{Quality cuts and orbital parameters}\label{quality cuts}
\annotatedtext{Describe age data, its uncertainty, and cut. Also, mention our results' robustness to age uncertainty.}{The main source of uncertainty in the MGAR arises from stellar age measurements. To minimize the contamination from binary evolution or mass transfer, we remove stars with masses $< 0.7\, M_\odot$ \citep{2023A&A...675A..26W}. Since the age estimates have a typical uncertainty floor of $\sim20\%$ \citep{2023A&A...675A..26W}, we exclude stars with age uncertainties exceeding 30\%. We have verified that applying a stricter 25\% cutoff yields similar results.}

Additional quality criteria include signal-to-noise ratio $\mathrm{SNR}>20$, vertical distance $|Z|<10\,\mathrm{kpc}$, and Galactocentric radius $5\,\mathrm{kpc} < R < 20\,\mathrm{kpc}$. \newemph{We require \texttt{flag\_feh\_apogee=0} and \texttt{flag\_afe\_apogee=0} to exclude stars with unreliable [Fe/H] or [$\alpha$/Fe] estimates outside the training grid.} \myemph{To avoid halo contamination, we retain only stars with $\mathrm{[Fe/H]}\geq-1.0$ dex and $v_\phi \geq 0$}. 

\newemph{We further exclude stars with relative uncertainty $>$ 40\% in any of the spatial or velocity parameters $X\in \{r, v_r, v_\phi, z, v_z \}$, i.e., we require}
$$
 \displaystyle\left|\dfrac{\Delta X}{X}\displaystyle\right|<40\% \text{ for all }X. 
$$
These quality cuts result in a clean sample of 246,671 stars.

\begin{figure}
\centering
\includegraphics[width=1\linewidth]{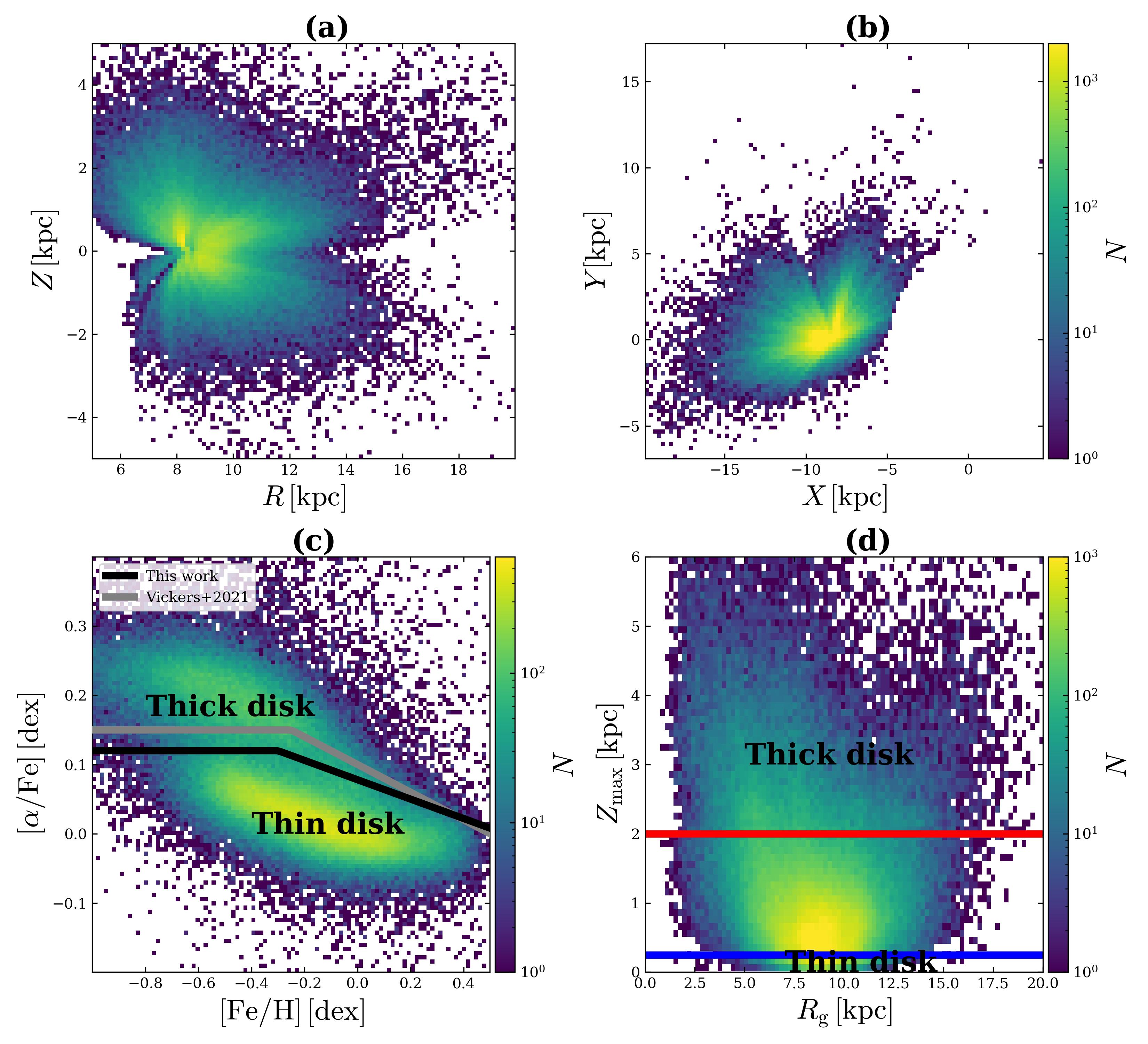}
\caption{Spatial and chemical distributions of the selected clean LAMOST DR8 sample. \textbf{(a)}: Galactic meridional distribution. \textbf{(b)}: Face-on view of the sample. \textbf{(c)}: Number density in the [$\alpha$/Fe]--[Fe/H] plane. The black line shows our chemical separation between the thin and thick disks ([$\alpha$/Fe] $= 0.12$ for [Fe/H] $< -0.3$, otherwise [$\alpha$/Fe] $= 0.12 - 0.14(\mathrm{[Fe/H]} + 0.3)$). The gray line indicates the V21 separation, which is not optimal for our sample. \textbf{(d)}: Number density in the $Z_\mathrm{max}$--$R_\mathrm{g}$ plane. The blue line at $Z_\mathrm{max}=0.25\,\mathrm{kpc}$ and red line at $Z_\mathrm{max}=2\,\mathrm{kpc}$ mark our geometric cuts for the thin and thick disk selections, respectively.}
\label{fig:data_dist}
\end{figure}

\annotatedtext{Explain how $R_\mathrm{g}$ and $Z_\mathrm{max}$ are calculated.}{To mitigate the impact of orbital blurring, we follow V21 and compute gradients using the Galactocentric radius of the guiding center $R_\mathrm{g}$. $R_\mathrm{g}$ and the maximum vertical excursion $Z_\mathrm{max}$ are calculated using the Galactic potential model \texttt{MWPotential2014}, implemented in the \texttt{galpy} package \citep{2015ApJS..216...29B}. We adopt the solar position $R_0 = 8.27\,\mathrm{kpc}$ and circular velocity $v_0 = 236\,\mathrm{km\, s^{-1}}$ \citep{2017MNRAS.472.3979S}. Figure~\ref{fig:data_dist} shows the distributions and selection criteria (next Section) used for defining thin/thick disk subsamples.}
\begin{figure}
\centering
\includegraphics[width=1\linewidth]{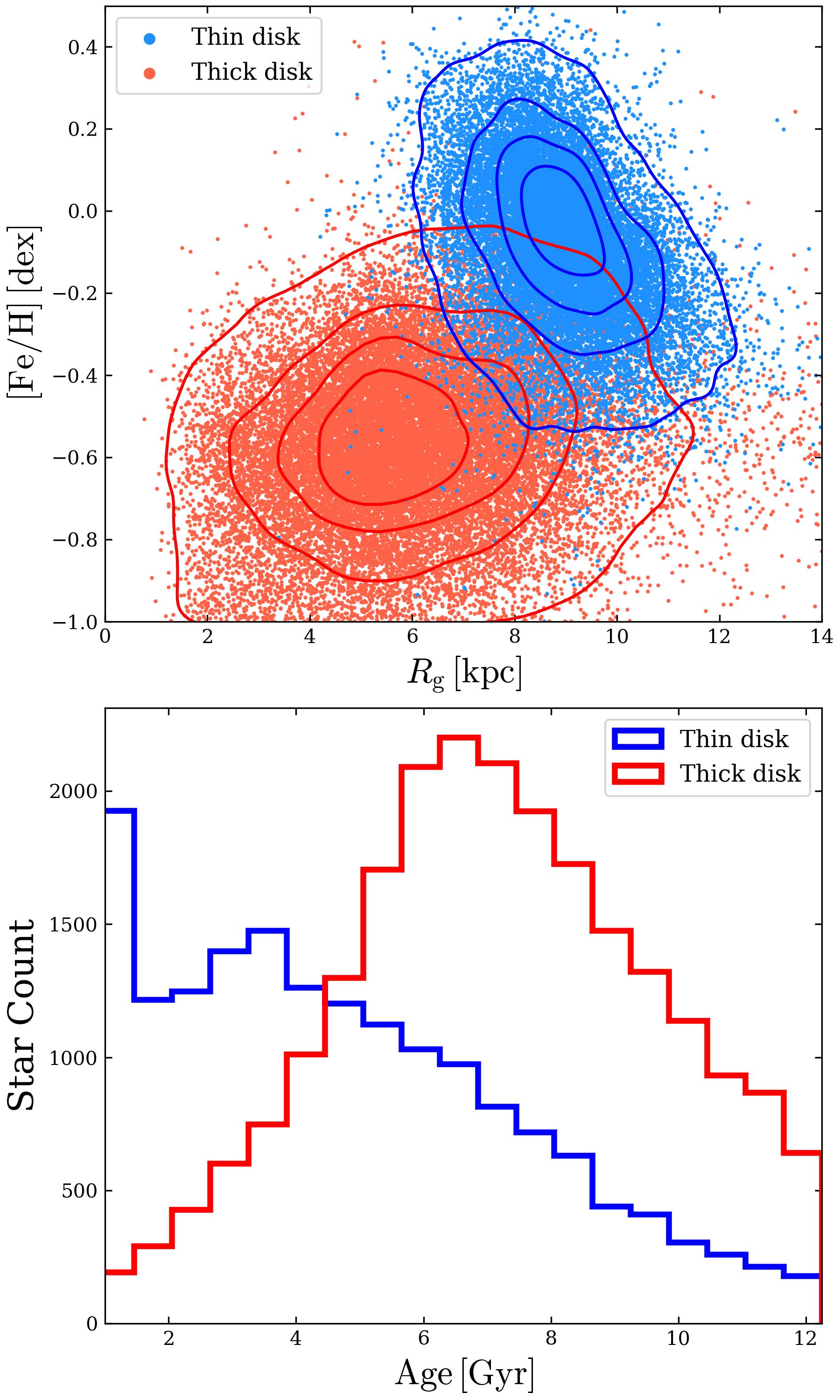}
\caption{[Fe/H]--$R_\mathrm{g}$ and age distributions for the selected thin (blue) and thick (red) disk stars. \textbf{Upper panel}: [Fe/H] vs. guiding center radius $R_\mathrm{g}$. Thin disk stars with near-solar metallicity show a gradient decreasing outward, while thick disk stars are more metal-poor (typical [Fe/H] $\sim-0.6$ dex) and exhibit a mildly positive gradient; \textbf{Lower panel}: Age distribution. Thin disk stars are generally younger than thick disk stars. The presence of old, metal-poor stars concentrated in the inner disk, and young, solar-metallicity stars in the outer disk, is consistent with inside-out disk formation.}
\label{fig:thin and thick}
\end{figure}
\subsection{Thin/thick disk subsamples}\label{subsamples}
\annotatedtext{Introduce two ways of selecting thin/thick disk.}{The MW’s disk can be separated into two components, either geometrically (via vertical excursion) or chemically (via [$\alpha$/Fe]–[Fe/H] bimodality). Geometrically, the MW's stellar density profile can be decomposed into two exponential components \citep[e.g.,][]{1983MNRAS.202.1025G}. In the $[\mathrm{\alpha/Fe}]$--$\mathrm{[Fe/H]}$ plane, there are two distinct sequences \citep[e.g.,][]{2015ApJ...808..132H, 2023ApJ...954..124I}; such a chemical bimodality reflects differences in nucleosynthetic timescales \citep{2013ARA&A..51..457N}. The $\alpha$-rich sequence is associated with rapid early star formation dominated by Type II supernovae, which produce more $\alpha$ elements (e.g., $^\text{16}$O, $^\text{20}$Ne, $^\text{24}$Mg, $^\text{28}$Si) relative to $^\text{56}$Fe. In contrast, Type Ia supernovae, with their longer delay times, enrich more Fe relative to $\alpha$ and contribute more to the $\alpha$-poor (solar-like) sequence \citep{2008gady.book.....B}.}%\textcolor{red}{[We are inclined to keep this part for completeness.]However, the gap between these two sequences indicates that different nucleosynthetic timescales are insufficient to explain the dichotomy. It is probable that additional processes, including mergers, gas accretion, and starbursts, are involved \citep{2023ApJ...954..124I}.}}

\annotatedtext{After introducing two selection ways, here we describe how we separate thin/thick disks.}{\triemph{As argued by V21, using chemical and geometric cuts together more effectively isolates the flattening MGAR trend than applying either cut alone. We combine both criteria to select ``pure'' thin and thick disk populations as follows}:
\begin{enumerate}
    \item Thin disk: [$\alpha$/Fe] below the black line in Figure~\ref{fig:data_dist}(c) and $Z_\mathrm{max}<0.25\,\mathrm{kpc}$
    \item Thick disk: [$\alpha$/Fe] above the black line in Figure~\ref{fig:data_dist}(c) and $Z_\mathrm{max}>2\,\mathrm{kpc}$.
\end{enumerate}}
\newemph{This yields 18,363 thin disk stars and 24,210 thick disk stars (7.4\% and 9.8\% of the total clean sample, respectively). Their distributions in the [$\alpha$/Fe]--[Fe/H] and $Z_\mathrm{max}$--$R_\mathrm{g}$ planes are shown in Section \ref{section:Metallicity Gradient}.}

\annotatedtext{Describe Figure~\ref{fig:thin and thick}}{Figure~\ref{fig:thin and thick} presents the [Fe/H]--$R_\mathrm{g}$ and age distributions of the selected subsamples. The thin disk shows a clear negative metallicity gradient centered around solar metallicity. In contrast, the thick disk is more metal-poor (typical $\mathrm{[Fe/H]}\sim -0.6 \,\mathrm{dex}$) and exhibits a slightly positive gradient. Thick disk stars are also more centrally concentrated and older, consistent with the inside-out formation scenario.}

\begin{figure}
\centering
\includegraphics[width=1\linewidth]{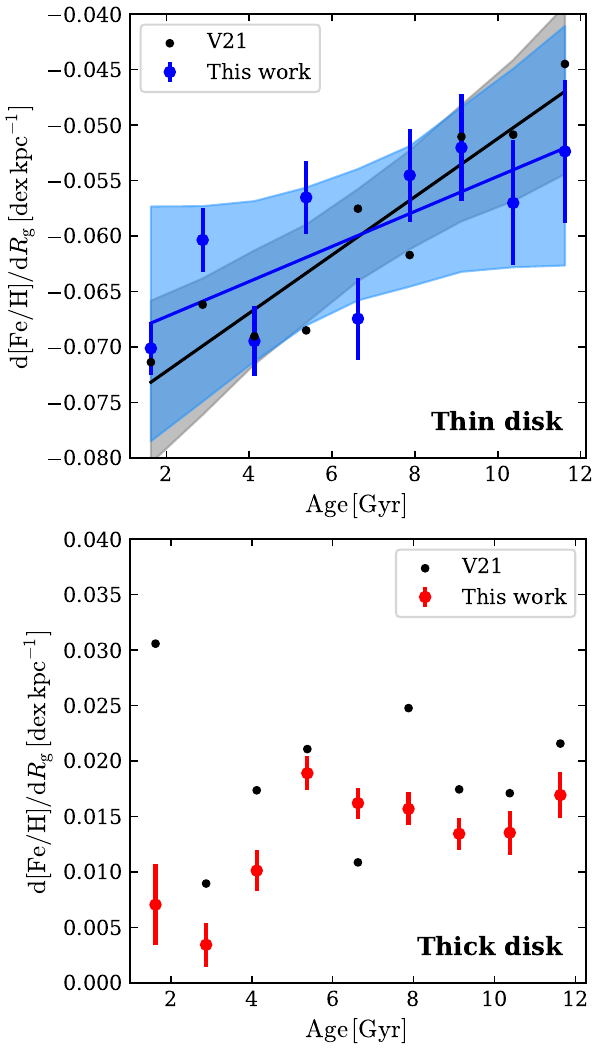}
\caption{Metallicity gradient–age relation (MGAR) in the thin (top) and thick (bottom) disks, compared to the results of V21 (black dots). Error bars are derived from the covariance matrix of the weighted OLS fits. \textbf{Upper panel}: MGAR in the thin disk. The blue line shows our fitted trend, with a flattening rate of $0.0016\,\mathrm{dex \, kpc^{-1}\, Gyr^{-1}}$; the shaded region denotes the 98\% highest density interval (HDI). The black line and gray shade show the fit for V21. Check the caption of Figure~\ref{fig:all_uncertainties} in the Appendix for prior specifications. \textbf{Lower panel}: MGAR in the thick disk. Our results yield a mean positive gradient of $\sim0.013\,\gradunit$.}
\label{fig:result}
\end{figure}
\subsection{Calculation of the MGAR}
\label{section:grad calc}
\annotatedtext{Describe the algorithm.}{We calculate the metallicity gradient $\mathrm{d[Fe/H]}/\mathrm{d}R_\mathrm{g}$ across nine equal-width age bins from 1 to 12.25 Gyr. Following V21, we apply the weighted ordinary least squares (OLS) regression from \texttt{SciPy} \citep{2020SciPy-NMeth} using weights $w_i=1/\Delta \mathrm{[Fe/H]}^2_i$, where $\Delta \mathrm{[Fe/H]}$ is the absolute metallicity uncertainty. The uncertainty in the fitted gradient is derived from the covariance matrix. 

To evaluate the effect of uncertainties in [Fe/H], $R_\mathrm{g}$, and age simultaneously, we perform an independent Bayesian fit in Appendix \ref{section:uncertainty} using the \texttt{PyMC} package \citep{pymc}. This test incorporates all three sources of uncertainty and confirms that the MGAR trends in both disks are robust.}

\section{Results}
\label{section:Metallicity Gradient}
\begin{figure*}
\centering
\includegraphics[width=1\linewidth]{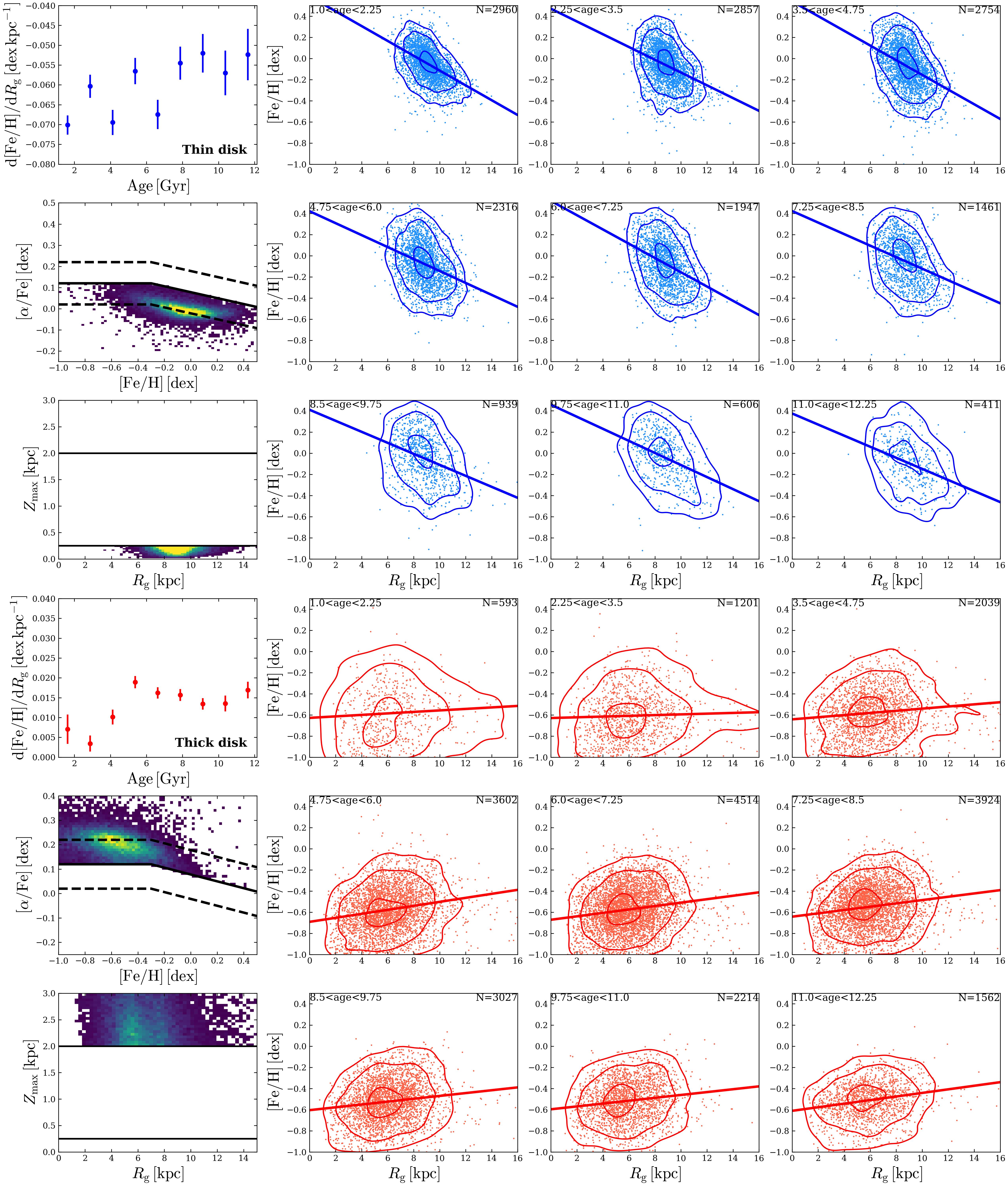}
\caption{Metallicity gradient fits for mono-age populations in the thin (top three rows) and thick (bottom three rows) disks. Each panel in the right three columns shows the [Fe/H]--$R_\mathrm{g}$ distribution for a specific age bin, with 20\%, 80\%, and 95\% contours and the size of the mono-age population.
The first column displays the MGARs, [$\alpha$/Fe]--[Fe/H], and $Z_\mathrm{max}$--$R_\mathrm{g}$ distributions for the two disk populations. \myemph{Notably, spurred contours appear in the fourth row at $R_\mathrm{g}>10\,\mathrm{kpc}$, which tend to suppress the positive gradient. After removing these stars, the thick disk gradient increases from $\sim 0.013\,\gradunit$ to $\sim 0.02\,\gradunit$ with reduced age-dependent fluctuation.}}
\label{fig:detail}
\end{figure*}

\annotatedtext{Introduce Figure~\ref{fig:result} and \ref{fig:detail} for the main results.}{Figure~\ref{fig:result} presents the MGAR in the thin and thick disks, compared to the results of V21 (black points). Gradient fits for each mono-age population are shown in detail in Figure~\ref{fig:detail}. Larger uncertainties for the oldest thin disk stars and the youngest thick disk populations are primarily due to limited sample sizes (see the age histograms in Figure~\ref{fig:thin and thick}).}

\annotatedtext{Don't forget about the robustness test for the main results in the appendix.}{We also examine how the MGAR responds to variations in selection criteria and orbital integration configurations in Appendices \ref{section:cuts} and \ref{section: galacticsetup}. Similar to V21, Appendix \ref{section:cuts} shows that the flattening MGAR in the thin disk primarily originates from low-$Z_\mathrm{max}$ stars, \triemph{which may be due to the favored migration for stars on the vertically coldest orbits \citep{2012MNRAS.422.1363S, 2014ApJ...794..173V}}. For the intermediate $Z_\mathrm{max}$, we observe a slightly non-monotonic MGAR, similar to V21 and \citet{2023MNRAS.526.2141W}. Appendix \ref{section: galacticsetup} further demonstrates that the choice of Galactic potential $\Phi$ and solar circular velocity $v_0$ can affect the numerical values of MGAR in the thin disk, but the overall flattening trend remains robust.}

\subsection{Metallicity gradient in the thin disk}
\annotatedtext{Describe MGAR in the thin disk. Compared with V21.}{The metallicity gradient in the thin disk gradually flattens as stellar age increases and aligns broadly with V21 and \citet{2023MNRAS.526.2141W}.
%at a rate of approximately $0.0016\,\mathrm{dex \, kpc^{-1}\, Gyr^{-1}}$.
%, although we find a slightly slower flattening rate ($0.003\,\mathrm{dex\, kpc^{-1}\, Gyr^{-1}}$ in V21). 
The gradient of the youngest stars reaches $-0.07\,\gradunit$ in agreement with V21 \triemph{and other studies} \citep{2020MNRAS.495.2673C, 2021ApJ...919...52Z, 2023A&A...674A.129W, 2023MNRAS.526.2141W, 2023MNRAS.525.2208R, 2024ApJS..272....8S}.}

\annotatedtext{Other MGAR in the literature.}{A comprehensive comparison of thin disk MGAR measurements using various stellar tracers is given in \citet{2023A&A...674A.129W}. Their Figure 5 highlights a distinctively V-shaped MGAR trend \triemph{with a dip around $5\sim 7 \,\mathrm{Gyr}$} in studies based on main-sequence turn-off (MSTO) stars \citep[e.g.,][]{2015RAA....15.1209X, 2019MNRAS.482.2189W}, which is absent in our result using RGB and RC stars. V-shapes with a dip around 2 Gyr have been reported using red giant stars from CoRoGEE \citep{2017A&A...600A..70A} and APOGEE \citep{2023A&A...678A.158A}. It is worth noting that the V-shapes seem to appear more frequently when the gradient is calculated with respect to $R_\mathrm{present}$ \citep{2015RAA....15.1209X, 2017A&A...600A..70A, 2019MNRAS.482.2189W} rather than $R_\mathrm{g}$ \citep[V21; ][]{2023MNRAS.526.2141W}, though the V-shape of MGAR in \citet{2023A&A...678A.158A} is not sensitive to either radius. The existence and characteristics of these V-shapes remain unclear and warrant further investigation with larger samples.}
\subsection{Metallicity gradient in the thick disk}
\annotatedtext{MGAR in the thick disk. Also, mention the anomalous flattening trend for young stars in the thick disk.}{The gradients in the thick disk are generally positive across all ages, with a mean value of $\sim 0.013\,\gradunit$. \revise{\citet{2017ApJ...850...25L} reported a higher value of $0.035\,\gradunit$ using 2,035 thick-disk giant stars from LAMOST DR3.} Recently, \citet{2023ApJ...950..142H} measured the metallicity gradient with respect to the angular momentum $L_z$ in their chemically thick (high-$\alpha$ metal-poor) disk: $1.2 \times  10^{-3}\,\mathrm{ dex\, kpc^{-1}\, km^{-1}\, s}$, which corresponds to $\mathrm{d[M/H]}/\mathrm{d}R_\mathrm{g} = 0.28\,\mathrm{dex \, kpc^{-1}}$ and is significantly larger than our findings here. \citet{2024ApJS..272....8S} found a positive gradient weaker than $\sim 0.01\,\mathrm{dex\, kpc^{-1}}$ in the high-$\alpha$ disk across different $|Z|$ bins.

Interestingly, our gradient increases steadily with decreasing stellar age, reaching a peak of $\sim0.019\,\gradunit$ at $\sim 5.4\,\mathrm{Gyr}$, but flattens at younger ages. This flattening is likely driven by a population of young, metal-poor thick disk stars. 
As shown in the fourth row of Figure~\ref{fig:detail}, \revise{the stars located at $R_\mathrm{g}>10\,\mathrm{kpc}$ help to suppress} gradients in the young thick disk panels. Our test confirms that thick disk stars at $R_\mathrm{g}>10\,\mathrm{kpc}$ exhibit a negative gradient ($\sim-0.02\,\gradunit$). Removing these stars reduces the flattening at young ages and raises the gradient values to $\sim0.02\,\gradunit$, but the overall MGAR trend remains. Overall, our results are consistent with V21, reaffirming the existence of a slightly positive MGAR in the thick disk.}

\section{Discussion}
\label{section:discussion}
\subsection{Origin of the flattening gradient in the thin disk}
As discussed in Section \ref{section:introduction}, radial migration (particularly churning) is a likely mechanism for redistributing stars after their formation, thereby contributing to the observed flattening of the MGAR (see Figure~\ref{fig:result}, upper panel). However, it remains uncertain whether the flattening primarily reflects stellar migration or intrinsic evolution in the ISM gradient over time.
\begin{figure}
\centering
\includegraphics[width=1\linewidth]{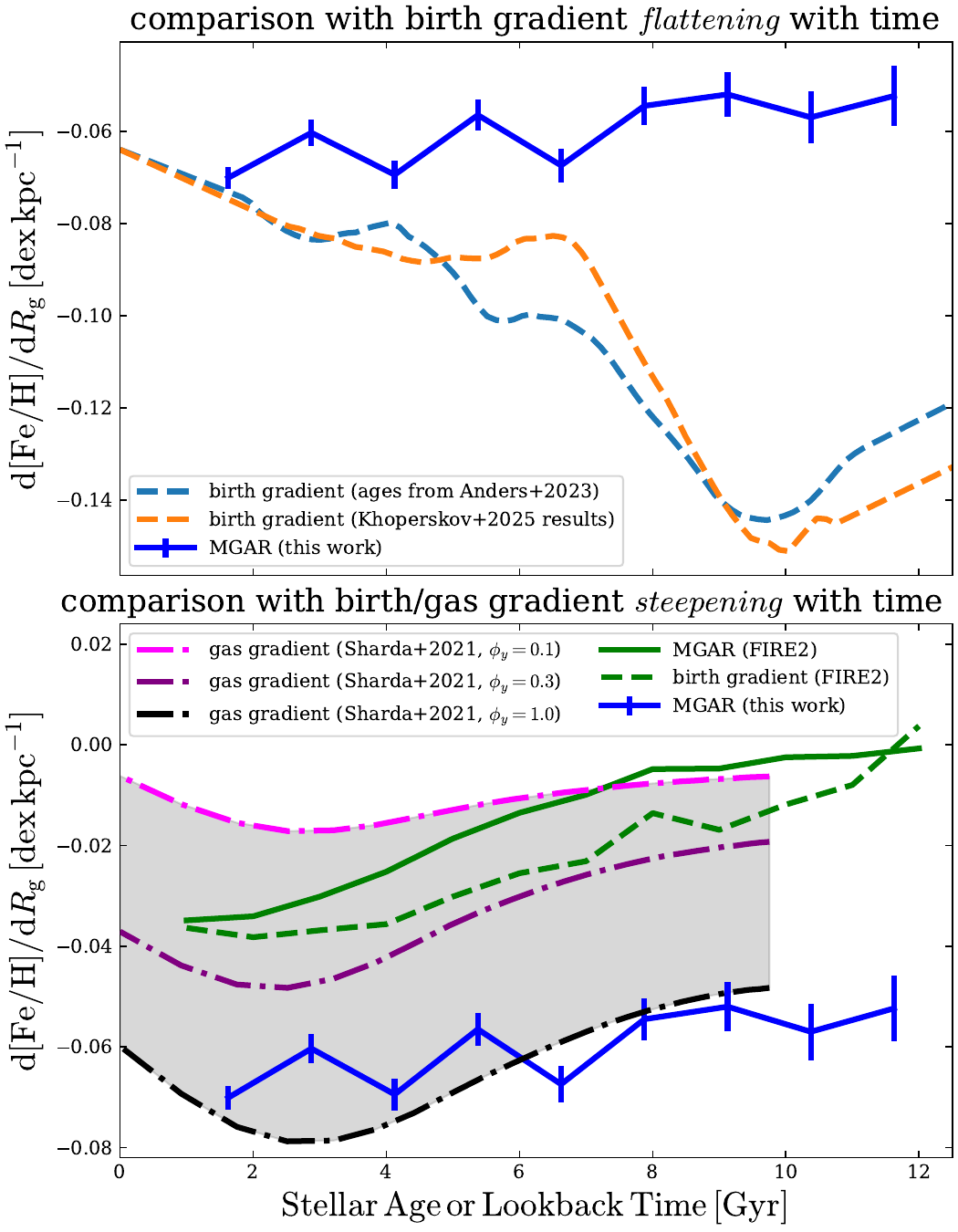}
\caption{Comparison between our observed MGAR in the thin disk and literature results \revise{(solid lines for MGARs, dashed/dash-dotted lines for stellar birth/gas-phase gradient evolutions)}. \textbf{Upper panel}: MGAR in the thin disk compared with reconstructed birth gradients \revise{retrieved from \texttt{Rbirth} python package (calculated from age data from \citet{2023A&A...678A.158A} and the selection function corrected results from \citet{2025A&A...700A..89K})}. The divergence between the MGAR and birth gradient at old ages may reflect the cumulative effect of churning. The gradient of the youngest stars aligns with the birth gradients (\revise{$\sim -0.065\,\gradunit$}), as little migration is expected to occur. \textbf{Lower panel}: MGAR compared with steepening gradients: \revise{the gas-phase metallicity gradient evolution model from \citet[dash-dotted lines and the grey shade. $\phi_y$ is the yield reduction factor of the model: larger $\phi_y$ for stronger ISM enrichment; smaller $\phi_y$ for metallicity losses through galactic winds.]{2021MNRAS.502.5935S}} and the birth gradient from FIRE2 simulations \citep{2024arXiv241021377G}, we also plot the MGAR from the FIRE2 simulations. See the text for further discussion.}
\label{fig:compare}
\end{figure}
Several studies have explored this question. \revise{Recent method for estimating $R_\mathrm{birth}$ introduced in Section \ref{section:introduction} found a gradually flattening of the birth metallicity gradients with time \citep{2025A&A...698A.267R}.} Similar flattening trends have been observed in external galaxies, where more massive systems tend to show shallower gradients. The analytic model of \citet{2021MNRAS.504...53S} captures this mass dependence, predicting that \revise{gas-phase metallicity }gradients measured in $\mathrm{dex\,kpc^{-1}}$ flatten with increasing stellar mass, while gradients in $\mathrm{dex\,r_e^{-1}}$ follow a U-shape ($r_e$ is the effective radius). Interestingly, applying this model to MW–like galaxies yields opposite gradient evolutions depending on assumptions: a \textit{steepening} ISM gradient over time \revise{in their Figure 10}, or a \textit{flattening} gradient \revise{in their Figure B1} if the gas velocity dispersion is \revise{reduced by $(1+z)$}, to account for possible overestimates in star formation rates \citep{2021MNRAS.502.5935S}. These discrepancies highlight the need for more robust theoretical treatments.

Recent FIRE2 simulations \triemph{challenged} the migration-driven scenario. \citet{2025ApJ...981...47G} found that the present-day MGAR closely follows the evolution of the stellar birth gradient (or ISM gradient then), suggesting minimal impact from radial migration. As shown in the simulations of \citet{2024arXiv241021377G}, gas-phase metals are efficiently mixed by turbulence and diffusion before star formation, suppressing spatial metallicity differences. As a result, prominent metallicity gradients emerge only after the disk reaches a settled state and turbulence diminishes. This interpretation emphasizes gas-phase processes over stellar redistribution.

\revise{Figure~\ref{fig:compare} contrasts our observed MGAR with several previously mentioned studies: solid lines for MGARs, dashed lines for stellar birth–gradient evolutions, and dot–dashed lines for gas–phase gradient evolutions.}
%We note that the gas–phase metallicity gradient ($\mathrm{d[O/H]/dR}$) is not identical to the stellar iron gradient ($\mathrm{d[Fe/H]/dR}$) measured in this work, so the comparison is qualitative.}

\subsubsection{Comparison with flattening birth gradient}

\revise{In the upper panel of Figure~\ref{fig:compare}, we compare our thin–disk MGAR with two reconstructed birth–gradient evolution trends retrieved from the \texttt{Rbirth} Python package, one based on the ages from \citet{2023A&A...678A.158A}, another from the selection–function–corrected results of \citet{2025A&A...700A..89K}. The upper panel reveals a growing divergence toward older ages (i.e., larger lookback time): the reconstructed birth gradients steepen substantially, whereas our observed MGAR becomes progressively flatter. As a result, the difference between the two reaches $\ge 0.08\,\gradunit$ by $\sim 10\,\mathrm{Gyr}$.}

\revise{This widening offset at old ages is most naturally interpreted as the cumulative effect of radial migration: stars born with a steep radial metallicity gradient can be redistributed over large radial ranges over several Gyrs, thereby flattening the present–day MGAR relative to the (much steeper) reconstructed birth gradients. Although the reconstructed curves also carry uncertainties and may be affected by systematics, the increased discrepancy toward older ages implies that migration is an important shaping mechanism for the oldest thin disk populations.}

\revise{We further note that \citet{2024MNRAS.535..392L} associated the sharp steepening in the reconstructed gradients around 10--12 Gyr with the Gaia–Enceladus/Sausage merger \citep{2018Natur.563...85H, 2018MNRAS.478..611B}. This event adds an additional complication when comparing reconstructed birth gradients with the present–day MGAR at large lookback times, but does not alter the qualitative conclusion that migration helps explain the observed divergence.}

\subsubsection{Comparison with steepening birth/gas gradient}

\revise{In the lower panel of Figure~\ref{fig:compare}, we compare the MGAR of our thin disk with both the MGAR and the birth gradients from FIRE2 simulations \citep{2024arXiv241021377G}, and with the analytical evolution of the gas–phase metallicity gradient from \citet{2021MNRAS.502.5935S}. In the analytical model, the parameter $\phi_y$ ($0\le\phi_y\le1$) represents the yield reduction factor: larger $\phi_y$ values imply that most supernova-produced metals remain in the ISM, while smaller $\phi_y$ values correspond to stronger metallicity losses through galactic winds.}

\revise{Our observed MGAR broadly follows the $\phi_y=1.0$ model for ages $\gtrsim3$ Gyr, except for an upturn at younger ages ($\lesssim 3$ Gyr), which is likely caused by dilution from late gas accretion. This agreement at intermediate and old ages suggests that, in this regime, the observed MGAR largely tracks the long–term evolution of the ISM metallicity gradient, and that radial migration plays only a secondary role. However, because the \citet{2021MNRAS.502.5935S} model assumes cylindrical symmetry and omits the bar and spiral structures, which are key drivers of radial migration in the Milky Way \citep{2002MNRAS.336..785S, 2010PhDT.......110R, 2025ApJ...983L..10Z, 2025A&A...701A..88M}, hence its predicted gas-phase gradients are likely steeper than in a fully realistic disk. Thus, the apparent agreement with the $\phi_y=1.0$ curve should be interpreted with caution.}
%(also note that the gas-phase metallicity gradient is not the same as the stellar metallicity gradient).}

\revise{For the FIRE2 simulations, both the MGAR and the birth gradients (green solid/dashed lines) lie between the $\phi_y=0.1$ and $\phi_y=0.3$ curves, substantially flatter than our observed MGAR, which is much closer to the $\phi_y=1.0$ curve. Thus, the vertical offset is plausibly explained by the stronger metal ejection in FIRE2, which consequently produces flatter ISM/birth gradients. Other possible contributors to the offset include uncertainties in distance measurements or the possibility that the real Milky Way experienced earlier disk settling than FIRE2 galaxies or nearby analogs \citep{2024arXiv241021377G}.}

\revise{Although the overall trends of MGAR and birth gradients in FIRE2 are similar (i.e., the agreement between the green solid line and green dashed line), which indicates limited radial migration within the simulations, the weak and short bars in FIRE2 \citep{2025ApJ...978...37A} suggest that the efficiency of churning may be underestimated relative to the real Milky Way. Therefore, the FIRE2 agreement between MGAR and the birth gradients reflects the underlying ISM evolution in the simulations rather than providing a direct constraint on the Milky Way’s migration efficiency.}

\subsection{Origin of the positive gradient in the thick disk}
The thick disk likely formed in the early evolutionary stages of the MW, when the disk was gas-rich and dynamically turbulent. Under such conditions, it is plausible that a weakly positive radial metallicity gradient could emerge. Simulations from the FIRE2 suite suggest that strong stellar feedback, especially from early starbursts, can drive metal-enriched outflows into the outer disk, leading to such positive gradients \citep{2025ApJ...986..179S}. \triemph{The accretion of pristine gas into the MW center during early epochs may also contribute to forming a positive metallicity gradient \citep{2010Natur.467..811C, 2012A&A...545A.133C}.} Recent JWST observations show positive metallicity gradient ($0.165\,\gradunit$) at z $\sim3$ \citep{2022ApJ...938L..16W}. For even higher redshifts, \citet{2024A&A...691A..19V} measured slightly positive gradients ($-0.03\sim0.14\,\gradunit$) in $6<z<8$, while \citet{2025ApJS..280...62L} reported flattening negative gradients ($-0.53\,\gradunit$ to $-0.34\,\gradunit$) from z $\approx 7$ to z $\approx 6$ (their Figure 2) and suggested that the slightly positive gradients in \citet{2024A&A...691A..19V} are resulted from merging systems.

Given the relatively small positive gradient we observe ($\sim0.013\,\gradunit$), another possibility is that the thick disk initially had no radial gradient. This would align with the clumpy formation scenario of the thick disk, where giant clumps heat the galactic disk and violent perturbations erase chemodynamical relations during the formation of the thick disk \citep[e.g.,][]{2014MNRAS.441..243I}.

%Given the relatively small positive gradient observed in our thick disk sample (mean value: $0.013\,\gradunit$), another plausible scenario is that the thick disk initially lacked a gradient altogether. This would be consistent with models in which the thick disk formed through dynamical heating and vertical thickening of an earlier clumpy, gas-rich disk \citep{2009ApJ...707L...1B}.

The flattening gradient with increasing age for thick disk stars older than $\sim5.4\,\mathrm{Gyr}$ might reflect the cumulative effects of radial migration within the thick disk \citep{2012MNRAS.422.1363S, 2018A&A...616A..86H}. Early turbulence and mixing could have dispersed stars from their birth radii, weakening any original gradient and leading to the flatter distributions seen today.

\triemph{Future studies could \revise{aim to correct for the selection function and} apply parameterized models to improve the robustness of MGAR measurements. Such parameterized models and chemical evolution models mentioned in Section \ref{section:introduction} may also help to distinguish between different birth gradient evolution scenarios shown in Figure~\ref{fig:compare}}.

\section{Conclusions}
\label{section:conclusion}

\annotatedtext{What we've done, what we get.}{Following the methodology of \citet[hereafter V21]{2021ApJ...922..189V}, we select chemically and kinematically ``pure'' thin and thick disk stars from LAMOST DR8 and compute the radial metallicity gradient ($\mathrm{d[Fe/H]}/\mathrm{d}R_\mathrm{g}$) across mono-age populations to investigate the metallicity gradient–age relation (MGAR) in both disks.

For the thin disk, we confirm a monotonic flattening of the MGAR with increasing stellar age. The fitted flattening rate is $\sim 0.0016\,\mathrm{dex \, kpc^{-1}\, Gyr^{-1}}$, which is lower than the value reported by V21 ($0.003\,\mathrm{dex\, kpc^{-1}\, Gyr^{-1}}$). The youngest thin disk stars exhibit a gradient of $-0.07\,\gradunit$, consistent with V21 and other studies. In contrast, the thick disk shows a generally positive MGAR with a mean value of $\sim 0.013\,\gradunit$, also in agreement with V21. Notably, we observe a flattening of the gradient for young thick disk stars (1.0–4.75 Gyr), likely due to a population of young, metal-poor stars at large guiding radii.}
%Following the methodology of \citet[V21]{2021ApJ...922..189V}, we select ``pure'' thin and thick disk stars from LAMOST DR8 and compute the radial metallicity gradient $\mathrm{d[Fe/H]}/\mathrm{d}R_\mathrm{g}$ across mono-age populations to characterize the metallicity gradient–age relation (MGAR) in the Milky Way’s disk. In the thin disk, we confirm a monotonic flattening of the gradient with age, with a slope of $0.0016\,\mathrm{dex\,kpc^{-1}\,Gyr^{-1}}$, which is smaller than the value reported by V21 ($0.003\,\mathrm{dex\,kpc^{-1}\,Gyr^{-1}}$). The gradient for the youngest population reaches $-0.07\,\gradunit$, in agreement with V21 and other studies. In the thick disk, we find a generally positive MGAR with a mean gradient of $0.013\,\gradunit$, also consistent with V21. Notably, a significant flattening is observed among the youngest thick disk populations (1.0–4.75 Gyr), likely driven by a population of young, metal-poor stars.

These MGAR trends reflect distinct chemodynamical histories for the MW’s thin and thick disks. In particular, the thin disk MGAR offers a valuable diagnostic for distinguishing between different formation scenarios, such as the ``inside-out'' model with significant radial migration and the ``upside-down'' model with minimal migration. Future studies using larger stellar samples with more precise ages, alongside improved chemodynamical modeling, will be crucial for disentangling these processes and refining our understanding of the Galaxy’s formation history.

\section*{Acknowledgements}

\revise{We thank the anonymous referee for helpful and constructive suggestions that strengthened the quality of this work.} We thank the developers and maintainers of the following software libraries, which were used in this work: \texttt{ArviZ} \citep{arviz_2019}, \texttt{Astropy} \citep{2022ApJ...935..167A}, \texttt{galpy} \citep{2015ApJS..216...29B}, \texttt{matplotlib} \citep{Hunter:2007}, \texttt{NumPy} \citep{harris2020array}, \texttt{pandas} \citep{reback2020pandas, mckinney-proc-scipy-2010}, \texttt{PyMC} \citep{pymc}, \texttt{seaborn} \citep{Waskom2021}, and \texttt{scipy} \citep{2020SciPy-NMeth}.

The research presented here is partially supported by the National Natural Science Foundation of China under grant Nos. 12533004, 12025302, 11773052; by China Manned Space Program with grant no. CMS-CSST-2025-A11; and by the “111” Project of the Ministry of Education of China under grant No. B20019. This work has made use of the Gravity Supercomputer at the Department of Astronomy, Shanghai Jiao Tong University.

%%%%%%%%%%%%%%%%%%%%%%%%%%%%%%%%%%%%%%%%%%%%%%%%%%
\section*{Data Availability}

The LAMOST DR8 astrometric and spectroscopic data, as well as the RGB and RC age catalog, are publicly available at: \url{http://www.lamost.org/dr8/v1.0/doc/vac}.



\bibliographystyle{mnras}
\bibliography{export-bibtex} % if your bibtex file is called example.bib

%%%%%%%%%%%%%%%%%%%%%%%%%%%%%%%%%%%%%%%%%%%%%%%%%%

%%%%%%%%%%%%%%%%% APPENDICES %%%%%%%%%%%%%%%%%%%%%

\appendix

\counterwithin{figure}{section}
\counterwithin{table}{section}
\renewcommand\thefigure{\thesection\arabic{figure}}
\renewcommand\thetable {\thesection\arabic{table}}
\newpage
Here, we present how our results vary with different fitting methods, thin/thick selections, and galactic setups.
\section{Fitting MGAR Incorporating All Uncertainties}
\label{section:uncertainty}
\begin{figure}
    \centering
    \includegraphics[width=1\linewidth]{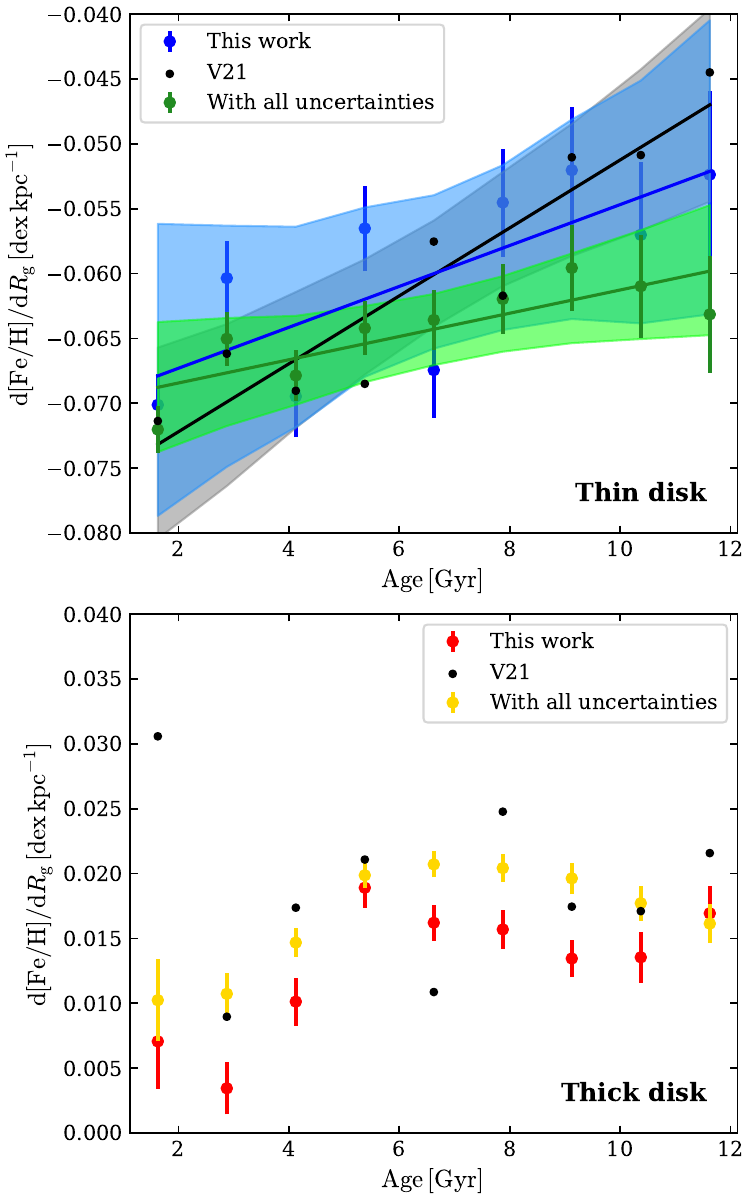}
    \caption{MGAR results from a Bayesian model that incorporates uncertainties in $\mathrm{[Fe/H]}$, guiding radius $R_\mathrm{g}$, and stellar age $\tau$. \textbf{Left panel}: MGARs in the thin disk. Blue points show the fiducial OLS results (this work), black points show V21, and forest green points are from the Bayesian model. Shaded regions denote the 98\% highest density intervals (HDI) for each fitted line. All fits adopt the same prior: slope $\sim \mathcal{N}(0, 2^2)$, intercept $\sim \mathcal{N}(0, 2^2)$. Solid line fitting model: $\text{gradient} \sim \mathcal{N}(\text{intercept} + \text{slope} \times \text{age bin center}, \epsilon^2)$, where $\epsilon \sim \mathrm{HalfCauchy}(\beta=5)$. \textbf{Right panel: }MGARs in the thick disk: OLS fitting (red), V21 (black), and the Bayesian model (gold).}\label{fig:all_uncertainties}
\end{figure}
We assess how uncertainties in [Fe/H], guiding radius $R_\mathrm{g}$, and stellar age $\tau$ affect the derived MGAR by implementing a Bayesian fitting procedure using \texttt{PyMC} \citep{pymc}. The absolute [Fe/H] uncertainty ($\Delta\mathrm{[Fe/H]}$) is taken directly from the LAMOST DR8 catalog. The absolute uncertainty in $R_\mathrm{g}$ ($\Delta R_\mathrm{g}$) is estimated by propagating position and velocity errors through 100 Monte Carlo realizations. The absolute stellar age uncertainty ($\Delta \tau$) is from \citet{2023A&A...675A..26W}, which includes an intrinsic method error of $\sim$ 20\% (i.e., $\delta \tau \gtrsim 20\%$).

We model the relation $\mathrm{[Fe/H]} \sim \nabla\, R_\mathrm{g} + b$, where $\nabla$ is the metallicity gradient and $b$ is the intercept, assuming a truncated normal likelihood $\widetilde{\mathcal{N}}$:
\begin{equation}\label{eq}
    \mathrm{[Fe/H]}_\mathrm{obs} \sim \widetilde{\mathcal{N}}(\mathrm{[Fe/H]} \mid \nabla\, R_\mathrm{g} + b, \Sigma^2),
\end{equation}
with the observed $\mathrm{[Fe/H]}_\mathrm{obs}$ truncated to the range $[-1.0,\,0.5]$ dex.
\begin{enumerate}
    \item Uncertainty in $R_\mathrm{g}$: We treat $R_\mathrm{g}$ as a latent variable drawn from a truncated normal distribution:
\begin{equation}
    R_\mathrm{g} \sim \widetilde{\mathcal{N}}(R_\mathrm{g} \mid R_\mathrm{g,obs}, \Delta R_\mathrm{g}^2), \quad R_\mathrm{g} \in [0,\, 20]~\mathrm{kpc}.
\end{equation}
    \item Uncertainty in [Fe/H]: The total variance $\Sigma$ combines intrinsic scatter $\sigma$ and absolute [Fe/H] uncertainty:
\begin{equation}
    \Sigma^2 \propto \sigma^2 + \Delta \mathrm{[Fe/H]}^2.
\end{equation}
    \item Uncertainty in age: To account for $\Delta\tau$, we apply a weight based on the probability that the true age lies within the assigned age bin. Assuming a truncated normal distribution for the true age within $[0,\,13]\,\mathrm{Gyr}$, the effective variance $\Sigma$ in Equation \ref{eq} becomes:
\begin{equation}
    \Sigma^2 = \dfrac{\sigma^2 + \Delta \mathrm{[Fe/H]}^2}{\left[ \displaystyle\int_{\mathrm{bin}} \widetilde{\mathcal{N}}(\tau \mid \tau_\mathrm{obs}, \Delta \tau^2) \, \mathrm{d}\tau \right]^2}.
\end{equation}
    \item Priors: The model uses the following priors:
\begin{equation}
\nabla \sim \mathcal{N}(0, 0.5^2),\,
b \sim \mathcal{N}(0, 0.5^2),\,
\sigma \sim \mathrm{HalfNormal}(0, 0.4^2).
\end{equation}
\end{enumerate}

Posterior sampling is performed using the \texttt{PyMC} package. The uncertainty of the gradient $\nabla$ is the standard deviation of the posterior.

Figure~\ref{fig:all_uncertainties} compares the MGAR obtained from this Bayesian model to our fiducial (OLS) results and V21 results. We find that our Bayesian model reduces fluctuations in the MGAR for both the thin and thick disks. \newemph{For the thin disk, the Bayesian MGAR trend shows the lowest flattening rate.} In the thick disk, the Bayesian MGAR shifts upward, though the age-dependent pattern remains similar. These results confirm that the qualitative MGAR trends are robust when all key uncertainties are included.

\section{Tests of Chemical/$Z_\text{max}$ Cut}\label{section:cuts}
We examine how changes to the chemical ([$\alpha$/Fe]--[Fe/H]) and geometric (i.e., $Z_\mathrm{max}$) selection criteria affect the MGAR.

\begin{figure}
    \centering
    \includegraphics[width=1\linewidth]{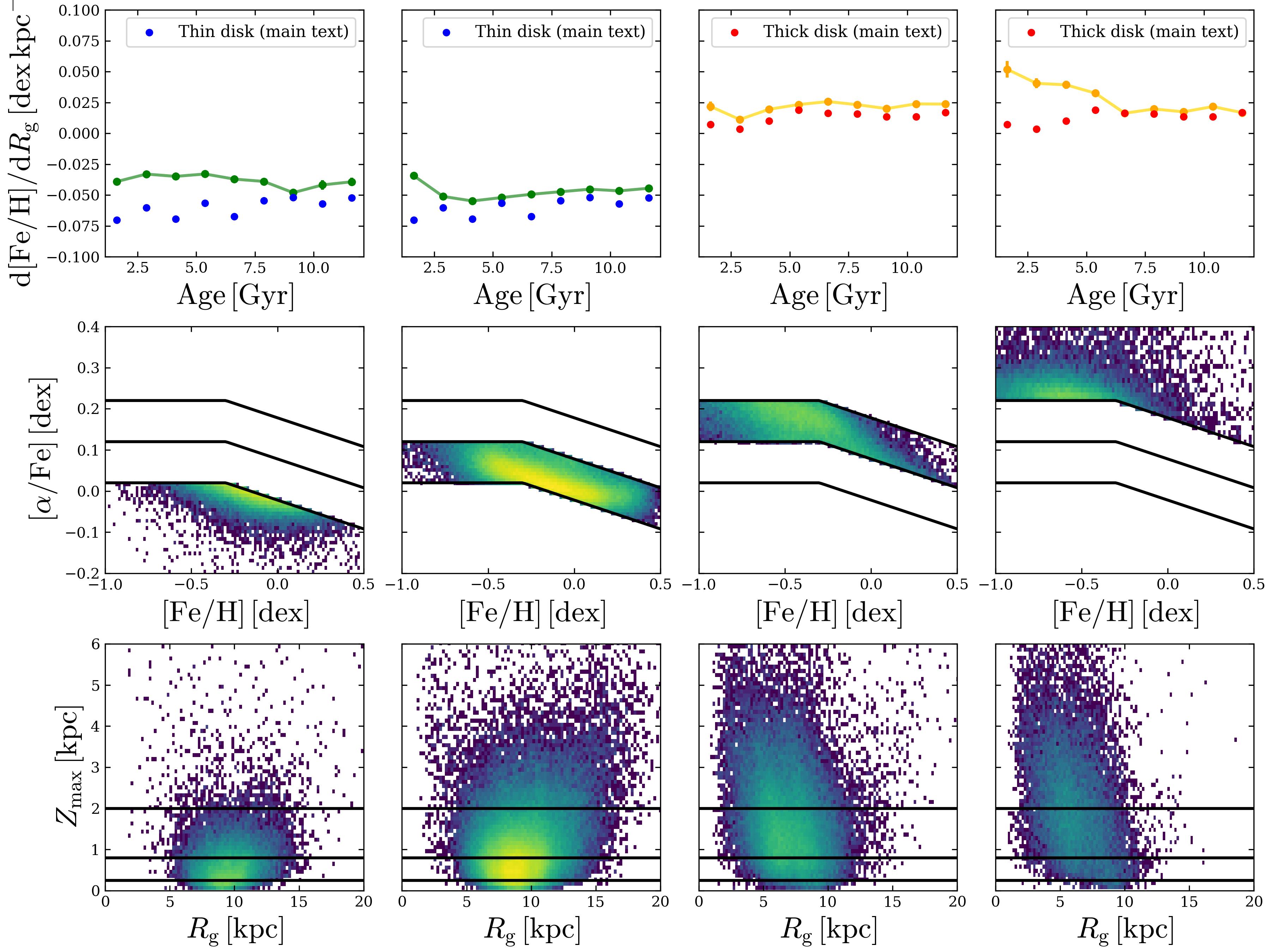}
  \caption{Effect of shifting the chemical division in the [$\alpha$/Fe]--[Fe/H] plane by $\pm0.1$ dex, blue and red dots are MGAR results from the main text. The top row displays the MGAR results without $Z_\mathrm{max}$ cut. The second row shows the chemical-only selection.}
    \label{fig:chemsubdiv}
\end{figure}

In Figure~\ref{fig:chemsubdiv}, we adjust the chemical division line in the [$\alpha$/Fe]--[Fe/H] plane by $\pm0.1\,\mathrm{dex}$ without applying the $Z_\mathrm{max}$ cut. The top row shows the resulting MGARs under these altered chemical selections.

\begin{figure}
    \centering
    \includegraphics[width=1\linewidth]{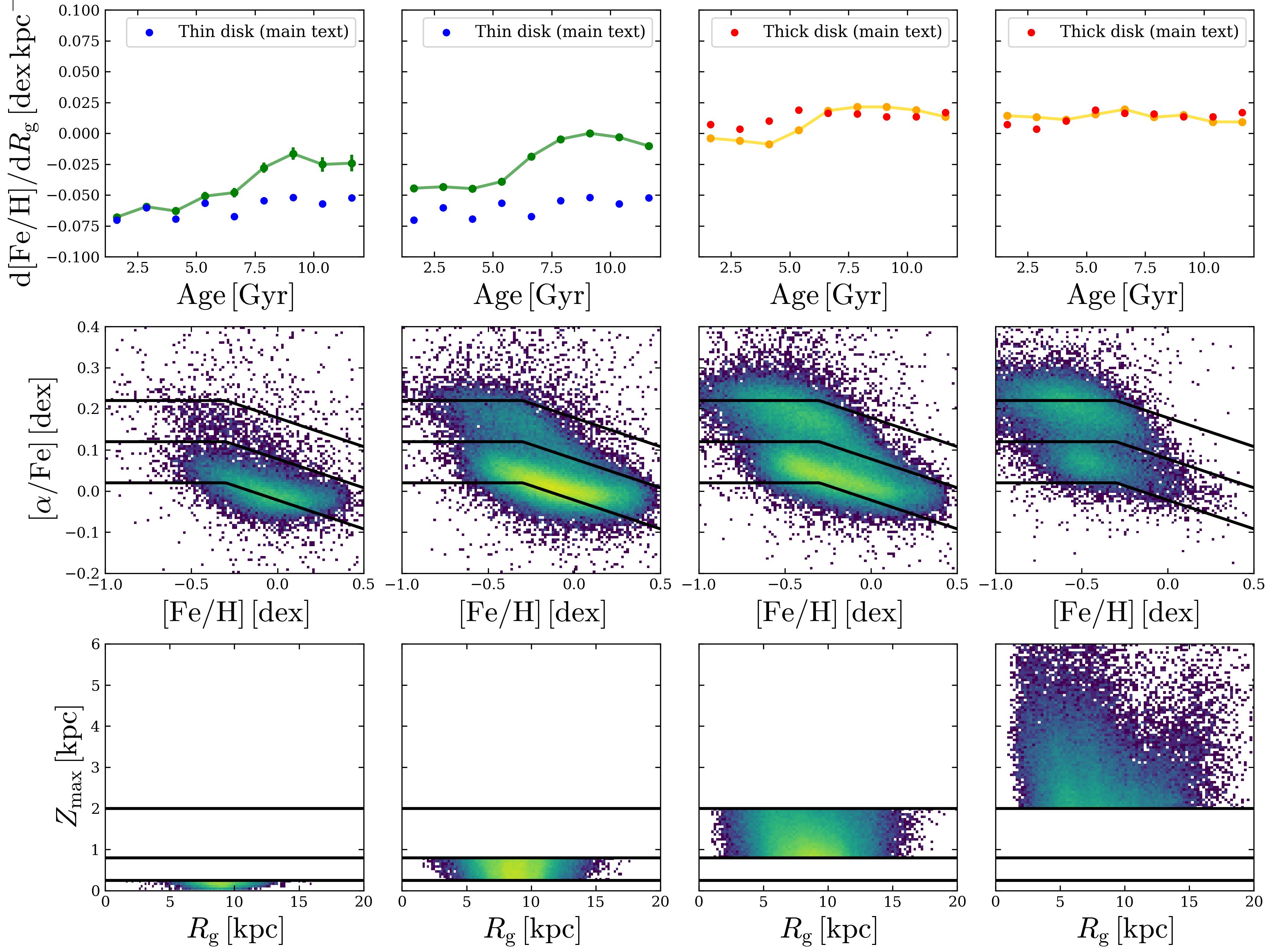}
  \caption{MGAR results for subsamples selected solely by $Z_\mathrm{max}$, with no chemical cut applied. The four columns correspond to $Z_\mathrm{max}$ = (0–0.25), (0.25–0.8), (0.8–2), and (2–6) kpc. The flattening MGAR trend in the thin disk is mainly contributed by stars with low $Z_\mathrm{max}$.}
    \label{fig:zmaxsubdiv}
\end{figure}

Figure~\ref{fig:zmaxsubdiv} explores the effect of purely geometric selections, dividing the sample into four $Z_\mathrm{max}$ bins: (0–0.25), (0.25–0.8), (0.8–2), and (2–6) kpc, with no chemical cut applied. The first and second columns of Figure~\ref{fig:zmaxsubdiv} show that the observed MGAR flattening in the thin disk mainly comes from stars with low $Z_\mathrm{max}$, \triemph{which may be due to the preferred migration for stars on the vertically coldest orbits \citep{2012MNRAS.422.1363S, 2014ApJ...794..173V}.} The slightly non-monotonic MGAR of stars with intermediate $Z_\mathrm{max}$ is also shown in V21 and \citet{2023MNRAS.526.2141W}.

\section{Sensitivity to Galactic Potential and Solar Parameters}
\label{section: galacticsetup}

\begin{figure}
    \centering
    \includegraphics[width=1\linewidth]{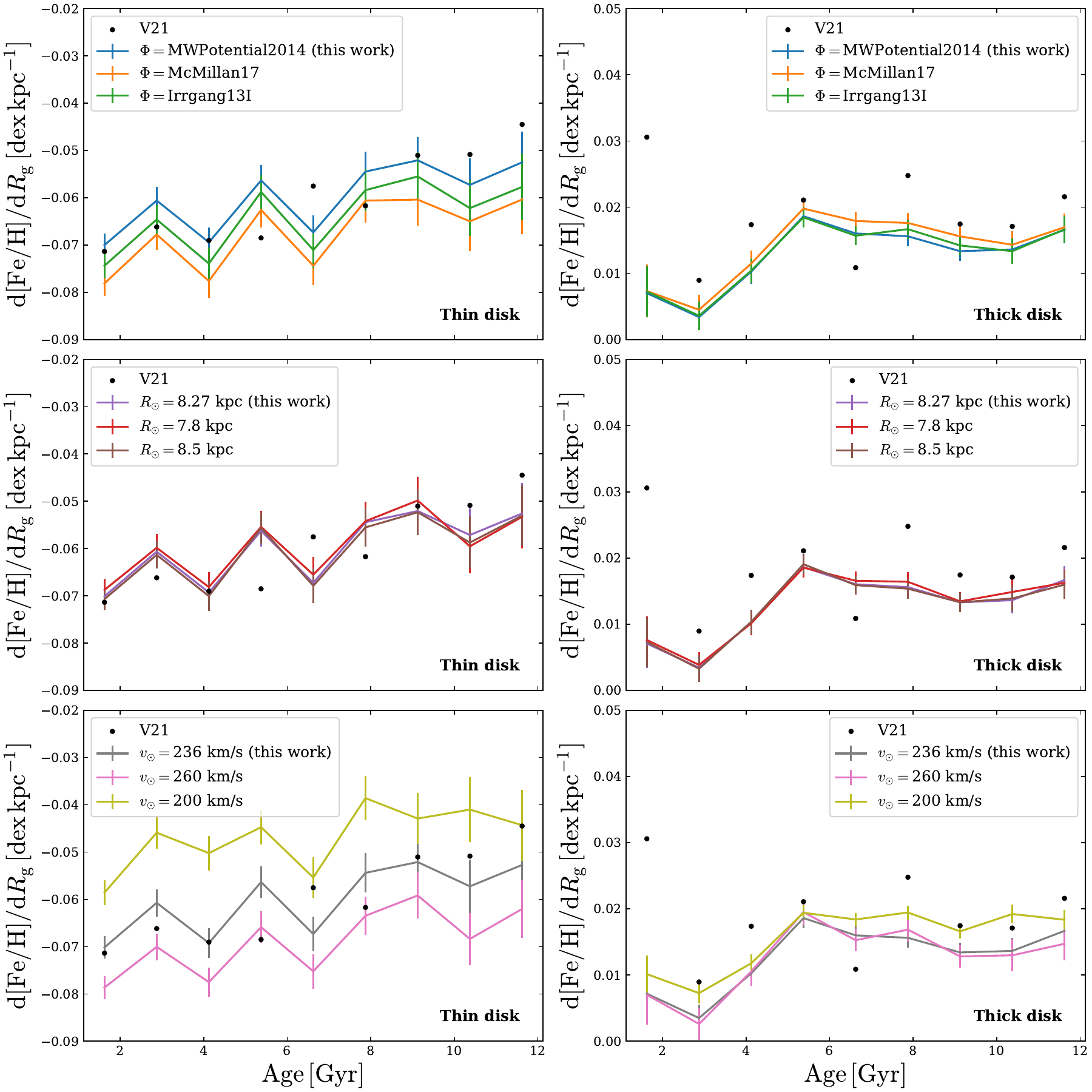}
  \caption{Robustness test of the MGAR under different Galactic configurations. The three rows correspond to variations in: (1) Galactic potential models used in \texttt{galpy} (\texttt{MWPotential2014}, \texttt{McMillan17}, \texttt{Irrgang13I}); (2) solar Galactocentric radius $R_0$ (8.27, 7.0, and 9.5 kpc); and (3) solar circular velocity $v_0$ (236, 260, and 200 km\,s$^{-1}$). The canonical setup in the main text adopts $\Phi=$\texttt{MWPotential2014}, $R_0 = 8.27\,\mathrm{kpc}$, and $v_0 = 236\,\mathrm{km\,s^{-1}}$. The thin disk MGAR is most sensitive to $\Phi$ and $v_0$.}
    \label{fig:discussion}
\end{figure}

The canonical results in the main text use $\Phi=$\texttt{MWPotential2014} as the Galactic potential, a solar Galactocentric radius of $R_0 = 8.27\,\mathrm{kpc}$, and a circular velocity of $v_0 = 236\,\mathrm{km\,s^{-1}}$ when calculating $R_\mathrm{g}$. To test the robustness of our results, we vary each of these parameters while keeping the others fixed, as shown in Figure~\ref{fig:discussion}. 

In the first row, we adopt three different Galactic potentials available in \texttt{galpy}: \texttt{MWPotential2014}, \texttt{McMillan17}, and \texttt{Irrgang13I}. In the second row, we vary the solar radius $R_0$ across 8.27, 7.0, and 9.5 kpc. In the third row, we test $v_0 =$ 236, 260, and 200 km\,s$^{-1}$. The resulting MGARs remain largely consistent across these configurations. We find that while the numerical values of the MGAR in the thin disk are sensitive to the adopted $\Phi$ and $v_0$, the overall MGAR remains robust across all panels in Figure~\ref{fig:discussion}.

%%%%%%%%%%%%%%%%%%%%%%%%%%%%%%%%%%%%%%%%%%%%%%%%%%

% Don't change these lines
\bsp	% typesetting comment
\label{lastpage}
\end{document}